# Seizure freedom after surgical resection of diffusion-weighted MRI abnormalities

Jonathan Horsley[1], Gerard Hall[1], Callum Simpson[1], Csaba Kozma[1], Rhys Thomas[2], Yujiang Wang[1,2,3], Jane de Tisi[3], Anna Miserocchi[3], Andrew McEvoy[3], Sjoerd Vos[3,4,5], Gavin Winston[3,6], John Duncan[3], Peter N. Taylor[1,2,3*]

1. CNNP Lab (www.cnnp-lab.com), Interdisciplinary Computing and Complex BioSystems Group, School of Computing, Newcastle University, Newcastle upon Tyne, United Kingdom
2. Translational and Clinical Research Institute, Faculty of Medical Sciences, Newcastle University, Newcastle upon Tyne, United Kingdom
3. Department of Clinical and Experimental Epilepsy, UCL Queen Square Institute of Neurology, University College London, London, United Kingdom
4. Centre for Microscopy, Characterisation, and Analysis, The University of Western Australia, Nedlands, Australia
5. Centre for Medical Image Computing, Computer Science Department, University College London, London, United Kingdom
6. Division of Neurology, Department of Medicine, Queen's University, Kingston, Canada

* Peter.Taylor@newcastle.ac.uk

# Abstract

## Importance

Many individuals with drug-resistant epilepsy continue to have seizures after resective surgery. Accurate identification of focal brain abnormalities is essential for successful neurosurgical intervention. Current clinical approaches to identify structural abnormalities for surgical targeting in epilepsy do not use diffusion-weighted MRI (dMRI), despite evidence that dMRI abnormalities are present in epilepsy and may relate to the epileptogenic zone.

## Objective

To investigate whether surgical resection of diffusion abnormalities relates to post-operative seizure freedom.

## Design, setting and participants

This retrospective case-control study was conducted between 2009 and 2022. Data were acquired at the National Hospital for Neurology and Neurosurgery, UK. Study participants included 200 individuals with drug-resistant focal epilepsy, who underwent resective surgery, and 97 healthy controls used as a normative baseline.

## Main Outcomes and Measures

Spatial overlap between diffusion abnormality clusters and surgical resection masks, and relation to post-surgical outcome.

## Results

Surgical resections overlapping with the largest abnormal cluster significantly correlated with sustained seizure freedom at 12 months (83% vs 55%; $p < 0.0001$) and over five years ($p < 0.0001$). Notably, resecting only a small proportion of the largest cluster was associated with better seizure outcomes than cases with no resection of this cluster ($p = 0.008$). Furthermore, sparing the largest cluster but resecting other large clusters still improved seizure freedom rates compared to no overlap ($p = 0.03$).

## Conclusions and Relevance

Mechanistically, our results suggest that abnormal clusters, identified using dMRI, are integral to the epileptogenic network, and even a partial removal of such an abnormal cluster is sufficient to achieve seizure freedom. The study highlights the potential of incorporating dMRI into pre-surgical planning to improve outcomes in focal epilepsy by reliably identifying and targeting diffusion abnormalities.

## Key Points

### Question

Does surgical resection of structural abnormalities, derived from diffusion-weighted MRI, relate to post-operative seizure freedom?

### Findings

Individuals who had diffusion-weighted MRI abnormalities resected in surgery were significantly more likely to be seizure-free at 12 months (83% vs 55%) and over five years.

### Meaning

Although not currently used clinically for this purpose, diffusion-weighted MRI may localise epileptogenic tissue. Prospective identification of these abnormalities may provide value for pre-surgical evaluation.

# Introduction

Up to half of people who undergo resective surgery for epilepsy continue to have seizures in the long term[1]. Current clinical approaches to determine which region(s) of the brain to remove in surgery involve the qualitative assessment of a variety of structural and functional data, including seizure semiology, structural MRI, EEG, fMRI, MEG, FDG-PET and SPECT[2]. These data are used to infer the location of the epileptogenic zone (EZ) - the area of the brain necessary for the generation of epileptic seizures[3]. By definition, in those people who continue to have seizures after surgery, the EZ was not sufficiently disrupted by the surgery, possibly due to mislocalisation. Successful surgery therefore requires accurate localisation and disruption of the EZ, and new data sources are needed to assist with this localisation in the clinic.

Many studies have shown that people with epilepsy have abnormalities detectable by diffusion-weighted MRI (dwMRI)[4,5]. Despite the magnitude of dMRI abnormalities being larger closer to the suspected epileptogenic zone in both temporal[4,6,7] and extratemporal[8] epilepsies, presurgical evaluations do not typically use dwMRI for localisation of the EZ. However, dwMRI may often be acquired to map white matter tracts and avoid postoperative neurological deficits caused by surgery[2,9–11]. Since dwMRI is already acquired in many cases, then it may provide additional benefit to existing pre-surgical evaluations at little extra cost, if it is shown to be able to localise the EZ.

It is currently not fully understood whether dwMRI abnormalities represent the EZ or some wider consequence of epileptic seizures. Previous work used dwMRI data to predict post-surgical outcomes, but often at group-level[12–15], with small sample sizes[13,14,16–20], or not validated using resection masks and post-surgical outcomes[21,22]. Additionally, many of these studies conducted analyses at the spatial scale of entire tracts or structural connectomes. White matter diffusion abnormalities, however, are not spread evenly along an entire tract, but may be localised to specific segments[15,23,24]. Some evidence suggested that the resection of these abnormal segments may be associated with seizure freedom, at least in specific tracts and types of epilepsy[15]. To improve clinical utility of dwMRI abnormalities, there is a need to a) identify localised abnormalities in individual subjects and b) determine whether resection of these abnormalities is associated with improved post-surgical outcomes.

In this study, we identified dwMRI abnormalities at the voxel level in 200 surgically treated people with drug-resistant focal epilepsy. We then assessed whether the location of these abnormalities were resected, and analysed the extent to which resection of dwMRI abnormalities resulted in seizure freedom after up to five years after surgery.

# Methods

## Subjects

We retrospectively studied 200 individuals with surgically treated drug-resistant focal epilepsy from the National Hospital of Neurology and Neurosurgery, London, United Kingdom. Analysis of pseudo-anonymised data from UCLH Epilepsy Surgery Database was approved by the Health Research Authority (22/SC/016).

Table 1 summarised the subjects in this cohort, stratified by post-surgical seizure freedom. Data from 97 healthy controls formed a normative comparison group. Of the 200 individuals with epilepsy, 83 had histopathological evidence of hippocampal sclerosis, 26 had focal cortical dysplasia, 25 had dysembryoplastic neuroepithelial tumor and 14 had cavernoma, 12 had dual pathology, two had glioma and one had treble pathology. The remaining 37 had some other unspecified pathology.

***Table 1: Patient data by 12 month post-surgical seizure freedom.*** *The difference in onset age between groups was assessed using a Wilcoxon rank-sum test. Other differences between groups were assessed using Chi-squared tests.*

| | ILAE 1,2 | ILAE 3+ | Test statistic |
|---|---|---|---|
| n | 139 | 61 | |
| Onset age, median (IQR) | 12 (14.5) | 15 (13.0) | $W = 3617, p = 0.10$ |
| Sex, male:female | 56:83 | 33:28 | $\chi^2 = 2.74, p = 0.10$ |
| Type, temporal:extratemporal | 108:31 | 47:14 | $\chi^2 \approx 0.00, p = 1$ |
| Side, left:right | 80:59 | 24:37 | $\chi^2 = 4.93, p = 0.03$ |
| MRI, non-lesional:lesional | 18:121 | 14:47 | $\chi^2 = 2.45, p = 0.12$ |

## Data acquisition

Diffusion-weighted MRI acquisition were obtained in two separate cohorts using different scanning protocols. The first cohort was collected between 2009 and 2013, and had 107 patients and 29 controls. The second cohort was collected between 2014 and 2019, and comprised 93 patients and 68 controls.

The first cohort of dwMRI data used a cardiac-triggered single-shot spin-echo planar imaging sequence[25] with echo time = $73ms$. Sixty contiguous 2.4 mm-thick axial slices were obtained covering the whole brain, with diffusion sensitizing gradients applied in each of 52 noncollinear directions (b value of 1,200 $s/mm^2$ [$\delta = 21ms$, $\Delta = 29ms$, using full gradient strength of $40mTm^{-1}$]) along with 6 non-diffusion-weighted scans. The gradient directions were calculated and ordered as described elsewhere[26]. The field of view was 24x24$cm$, and the acquisition matrix size was 96 × 96, zero filled to 128 × 128 during reconstruction, giving a reconstructed voxel size of 1.875 × 1.875 × 2.4 mm.

The second cohort of dwMRI data were acquired using a single-shot spin-echo planar imaging sequence with echo time = $74.1ms$. Seventy contiguous 2 mm-thick axial slices were obtained covering the whole brain. A total of 115 volumes were acquired with 11, 8, 32, and 64 gradient directions at b-values of 0, 300, 700, and 2500 $s/mm^2$ respectively ($\delta = 21.5ms$, $\Delta = 35.9ms$) as well as a single b = 0-image with reverse phase-encoding (B0). The field of view was 25.6x25.6$cm$, and the acquisition matrix size was 128 × 128, giving a reconstructed voxel size of 2 × 2 x 2 mm.

## Data processing and registration

The dMRI scans from both cohorts were processed identically. The scans were de-noised[27], Gibbs-unringed[28] and corrected for signal drift[29]. Furthermore, since one cohort did not have reverse phase-encoded B0s, the Synb0-DisCo[30,31] tool was used to create a non-distorted synthetic image from each participant’s corresponding T1 structural MRI. The Synb0-DisCo tool was run for both cohorts irrespective of the existence of reverse phase-encoding images to ensure continuity between the processing of the two cohorts. The calculated non-distorted synthetic image was subsequently input into TOPUP[32,33] and EDDY[34] to correct for warping, eddy current-induced distortions and motion. Lastly, correction for signal bias was applied using N4 bias field correction[35].

After pre-processing, tensor maps were calculated using FSL’s DTIFIT tool, and the FA maps from each individual were registered to a standard template (HCP_10065_FA) in MNI-152 standard space (Figure 1A). All b-values were used in tensor reconstruction for both cohorts. Registration used the *antsRegistrationSyN.sh* script from ANTs, which employed both linear (affine) and non-linear (diffeomorphic SyN) transformations[36,37]. Using the transformation files calculated from the prior registration, all tensor maps (FA, MD, AD and RD) were then moved into standard space using the *antsApplyTransforms* tool with a trilinear interpolation. No other smoothing was applied.

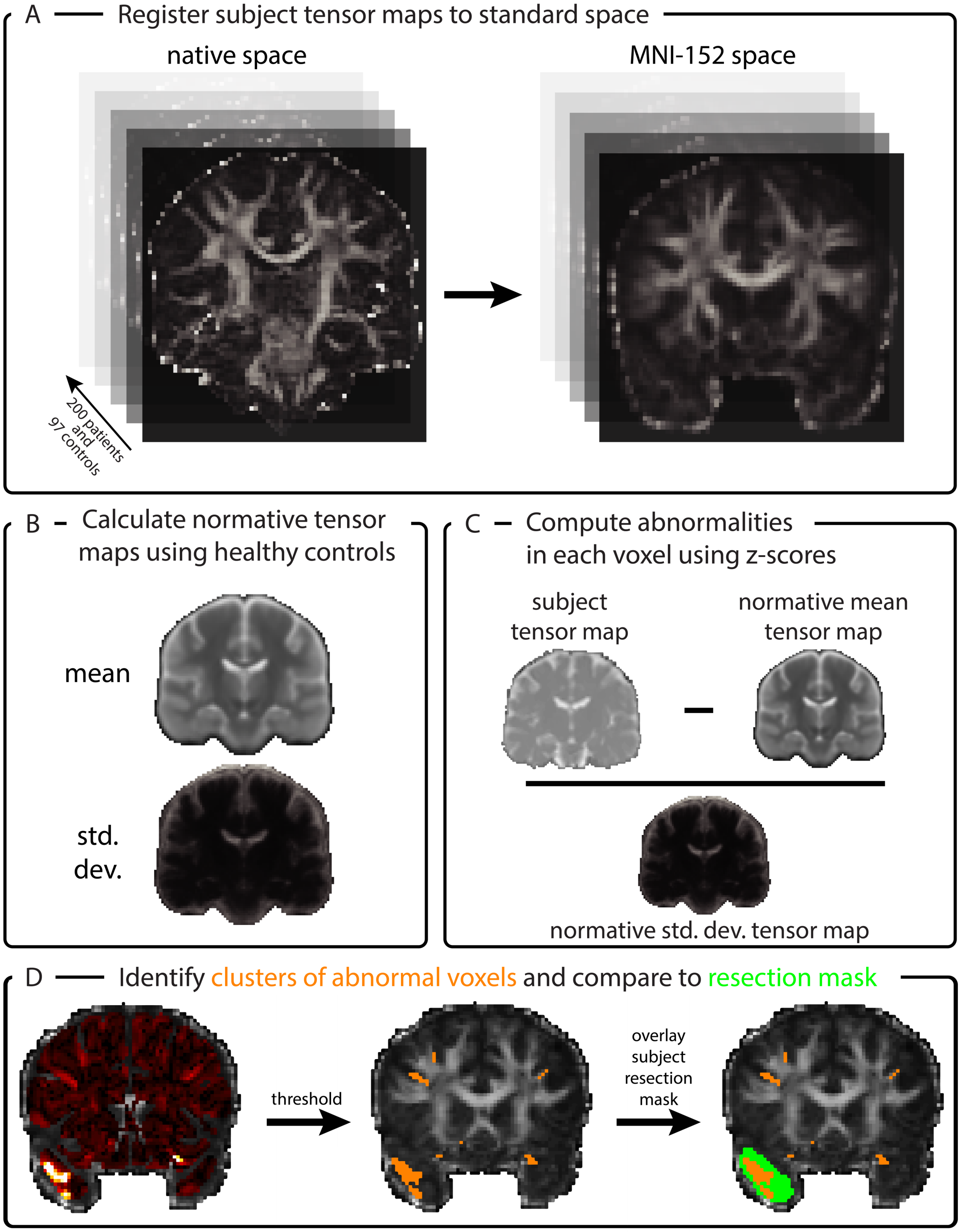

***Figure 1: Abnormality calculation pipeline**. A) All subjects were registered to the same MNI-152 standard space. B) Normative tensor maps of diffusion were calculated for MD, AD, RD and FA by calculating the mean and standard deviation of the tensor values across healthy*

*controls in each voxel (shown here for MD only). C) Abnormality values were calculated by z-scoring each subjects' tensor values against the same voxel in healthy controls. D) Abnormalities were thresholded at z = 3, and compared to the resection mask for each subject.*

## Abnormality calculation

All analysis was performed in R version 4.3.0.

Normative diffusion tensor maps were created using healthy controls. We computed both the mean and standard deviation of FA, MD, AD and RD in each voxel across controls (Figure 1B). Separate normative maps were created for both cohorts. Voxels in CSF, brain stem and cerebellum were discounted. These normative maps acted as a healthy baseline, against which we assessed individual subjects.

We calculated abnormalities in each voxel by z-scoring against the corresponding voxel in the normative map of the same cohort (Figure 1C). This z-scoring was done separately for FA, MD, AD and RD. In each case, the abnormality values in each voxel specified the number of standard deviations away from the healthy mean. Abnormalities were calculated independently for each voxel but abnormalities in neighbouring voxels may be more indicative of a true abnormal signal rather than noise. To boost the signal-to-noise ratio by considering neighbouring abnormalities, we applied probabilistic threshold free clustering enhancement (pTFCE)[38].

Abnormal voxels were defined as those exceeding a (pTFCE-enhanced) z-score of 3, and clusters of neighbouring abnormal voxels were identified (Figure 1D). Supplementary Analysis 1 shows results for alternative thresholds. Clusters were ordered by volume and substantially large clusters were detected. Specifically, we defined substantially large clusters in each individual as those that exceeded a volume threshold that differentiated them from smaller, potentially spurious clusters. This threshold was derived in a data-driven manner using change-point analysis[39] with a Gamma prior distribution.

## Resection mask generation

Resection masks were generated using a semi-automated approach as described previously[40]. Briefly, we used postoperative imaging to generate masks of the tissue that was subsequently resected. Masks were initially generated automatically using a custom built software pipeline, using FastSurfer[41], ANTs[36] and ATROPOS[42]. These automated masks were visually inspected and, if needed, manually corrected to ensure quality. The resection masks were then registered to the same standard (MNI-152) space as the abnormality maps.

Within each subject, cluster abnormality maps were overlaid with the resection mask. From this, we calculated the proportion of the largest cluster resected. The same proportion was calculated for all substantially large clusters within a subject. We compared these cluster resection proportions to the likelihood of a person remaining free from disabling seizures following surgery. We hypothesised that resection of the largest, or other substantially large clusters would be associated with post-operative seizure freedom.

# Results

Clusters of diffusion abnormalities were calculated in each subject for MD, which we will present in the following main text. Alternative measures (AD, RD and FA) for abnormality cluster analyses are presented in Supplementary Analysis 2 for completeness, and show broadly similar results.

## Resection of the largest cluster is associated with good outcome

First, we investigated whether resection of the largest cluster was associated with post-surgical seizure freedom using survival analysis. Within an individual, the largest cluster was defined as resected if there was any overlap with the resection mask. At each yearly follow-up, a person was defined as seizure-free if they had no debilitating seizures, i.e ILAE 1 or 2, otherwise they were defined as not-seizure-free.

After 12 months, the seizure freedom rate amongst those with the largest cluster resected was 83%, compared to 55% amongst those with the largest cluster spared. Whether or not the largest cluster was resected significantly predicted outcome in the long (5 years) term ($p < 0.0001$). Sample sizes at each year of follow-up are presented in Supplementary Table 1. This trend was observed in both MRI negative ($n = 32$; $p = 0.07$) and MRI positive cohorts ($n = 168$; $p = 0.002$; Supplementary Analysis 3), in both scanning protocol cohorts (Supplementary Analysis 4) in both TLE and ETLE cohorts (Supplementary Analysis 5) and in both left and right-sided resection cohorts (Supplementary Analysis 6).

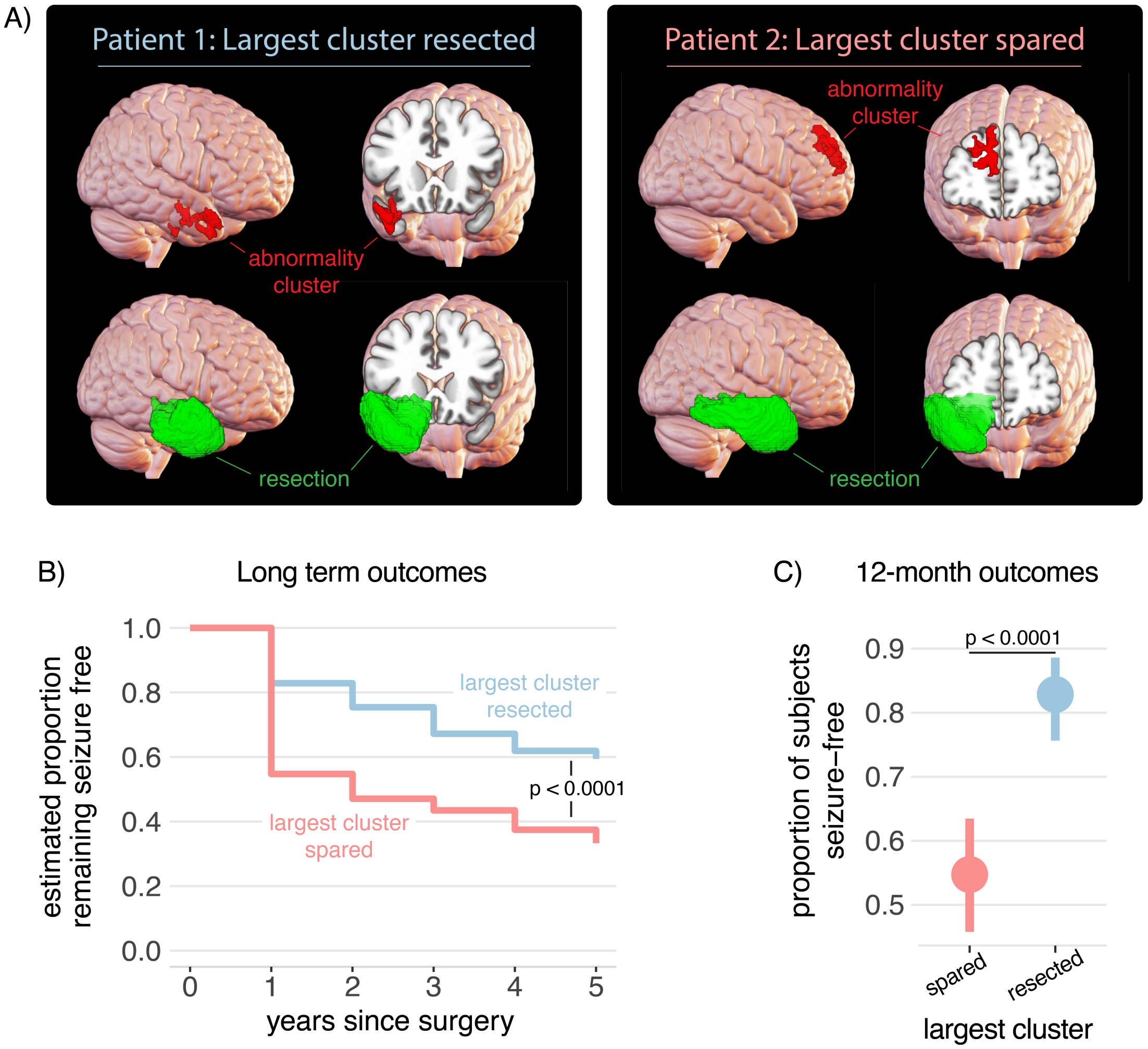

***Figure 2: Resection of the largest cluster is associated with good outcome**. A) The largest cluster of (MD) diffusion abnormalities are shown in red for two example subjects. In the left-hand subpanel (Patient 1), the resection mask (green) overlapped with the largest cluster. This subject was seizure-free (ILAE 1) for the three years of available follow-up data. In the right-hand subpanel (Patient 2), the resection mask (green) did not overlap with the largest cluster. Note: Patient 2 was clinically assessed as MR-negative. This subject was not seizure-free (ILAE 5) for the four years of available follow-up data. Across the cohort, overlap between the largest cluster and the resection mask was associated with a significantly improved rate of seizure freedom B) over five years and C) at 12 months post-surgery. Error bars indicate 90% confidence intervals.*

## Resection of even a small proportion of the largest cluster can still lead to a good outcome

In the previous analysis, if the largest cluster overlapped at all with the resection mask, the cluster was classed as resected. However, this did not answer whether the *amount of overlap* is important in predicting post-surgical seizure freedom. In this analysis, we only considered seizure freedom at 12 months to maximise the sample size. Of the 105 individuals with the largest cluster resected, a wide range in *proportion resected* was observed (Figure3B; median: 51%; IQR: 65%).

To investigate the importance of the proportion of the largest cluster resected on predicting outcome, we considered two subsets of the data. The first subset considered only those with a small proportion (no more than 30%) of the largest cluster resected ($n = 35$). Of these subjects, 28 (80%) were seizure-free at one year post-surgery. Despite minimal overlap between the largest cluster and the resection, these subjects were still significantly more likely to be seizure-free, compared to those with no overlap between the largest cluster and the resection ($p = 0.008$; Figure 3C).

The second subset considered only those with a large proportion (at least 70%) of the largest cluster resected ($n = 37$). Of these subjects, 34 (94%) were seizure-free at one year post-surgery. These subjects were significantly more likely to be seizure-free, compared to those with no overlap between the largest cluster and the resection ($p < 0.0001$; Figure 3C), but not significantly more likely to be seizure-free than those with a small proportion of the largest cluster resected ($p = 0.07$).

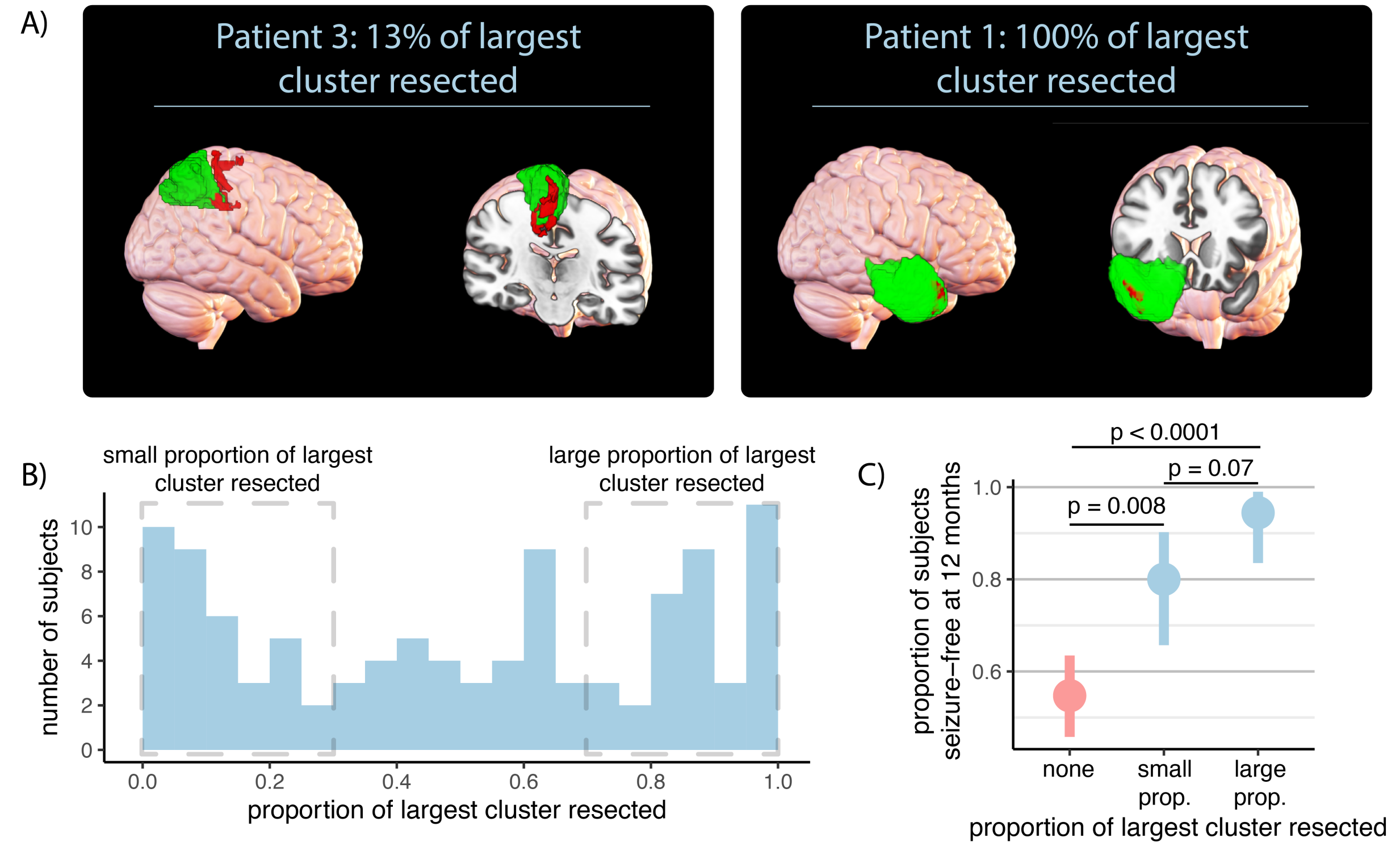

***Figure 3: Resection of even a small proportion of the largest cluster can still lead to a good outcome****. A) The proportion of the largest cluster resected are shown for two example subjects. Additional example subjects are shown in Supplementary Analysis 8. In the left-hand subpanel, a small proportion (13%) of the largest cluster was resected. In the right-hand subpanel, 100% of the largest cluster was resected. B) Histogram showing the proportion of the largest cluster resected, for those subjects with an overlap between the resection and the largest cluster. People who had no more than 30% (small proportion) or more than 70% (large proportion) of the largest cluster resected were identified. C) Resection of a small ($p = 0.008$) and large proportion ($p < 0.0001$) of the largest cluster were both associated with an improved chance of seizure freedom at one year post-surgery. Error bars indicate 90% confidence intervals.*

## Cluster number, volume, and distribution does not explain outcome

More than one abnormal cluster may exist in some patients. Other substantially large clusters were detected using change point analysis (Figure 4A). The number of substantial clusters detected across the cohort was between 1 and 11 (Figure 4B), but this did not differ between seizure-free and not-seizure-free subjects ($p = 0.30$). Similarly, the number of abnormal voxels contained within the substantial clusters did not differ between seizure-free and not-seizure-free subjects ($p = 0.84$).

For those subjects with several clusters detected, the clusters were often in multiple lobes, and no significant differences in the number of lobes affected were observed between seizure-free and not-seizure-free subjects (Supplementary Analysis 7).

Taken together, the number, total volume, and distribution of clusters were not related to outcome.

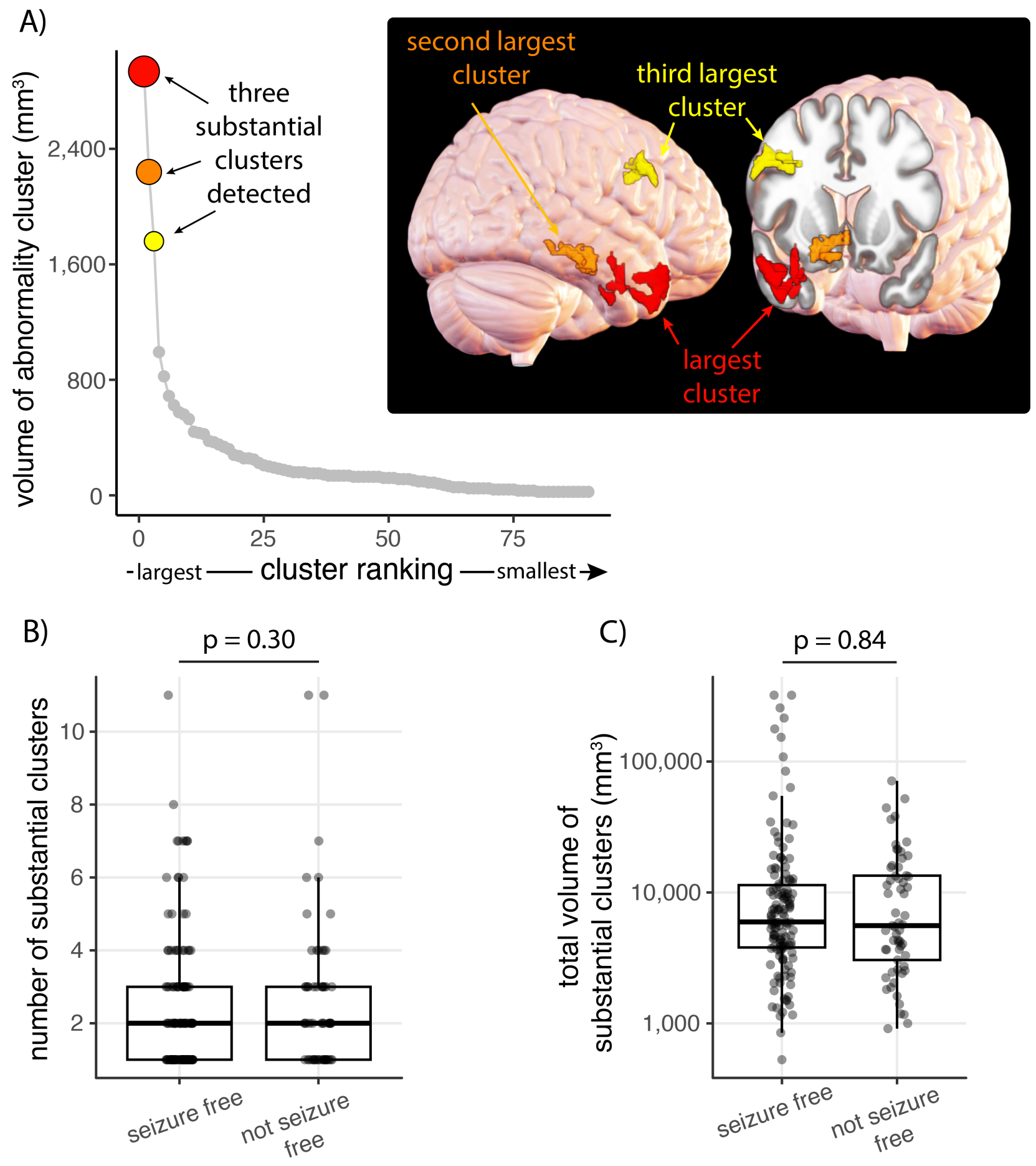

***Figure 4: Multiple substantial abnormality clusters may exist, but this does not explain outcome.*** *A) Within a subject, clusters were ranked by volume. The number of substantial clusters were detected using change-point analysis. In this example subject, three substantial clusters were detected. Neither B) the number of substantial clusters detected ($p = 0.30$) nor C) the volume of substantial clusters ($p = 0.84$) explained outcome at one year post-surgery.*

## Resection of other clusters can still lead to a good outcome

Next, we incorporated information about whether these substantially large clusters were resected into a survival analysis. We were particularly interested in considering the scenario in which the largest cluster was spared (Figure 5A). The possibilities were:

1. largest cluster was spared and no other substantial clusters were detected (n=22);
2. largest cluster was spared and at least one other substantial cluster was resected (n=35);
3. largest cluster and all other substantial clusters were spared (n=38).

The highest rates of seizure freedom occurred when the largest cluster was resected (83% seizure free at 1 year, Figure 5B), and the lowest rates of seizure freedom occurred when multiple substantial clusters existed but none were resected (47% seizure free at one year). In the case where the largest cluster was spared, but other substantially large clusters existed, resection of at least one of these clusters significantly improved the probability of seizure freedom ($p = 0.03$, Figure 5C).

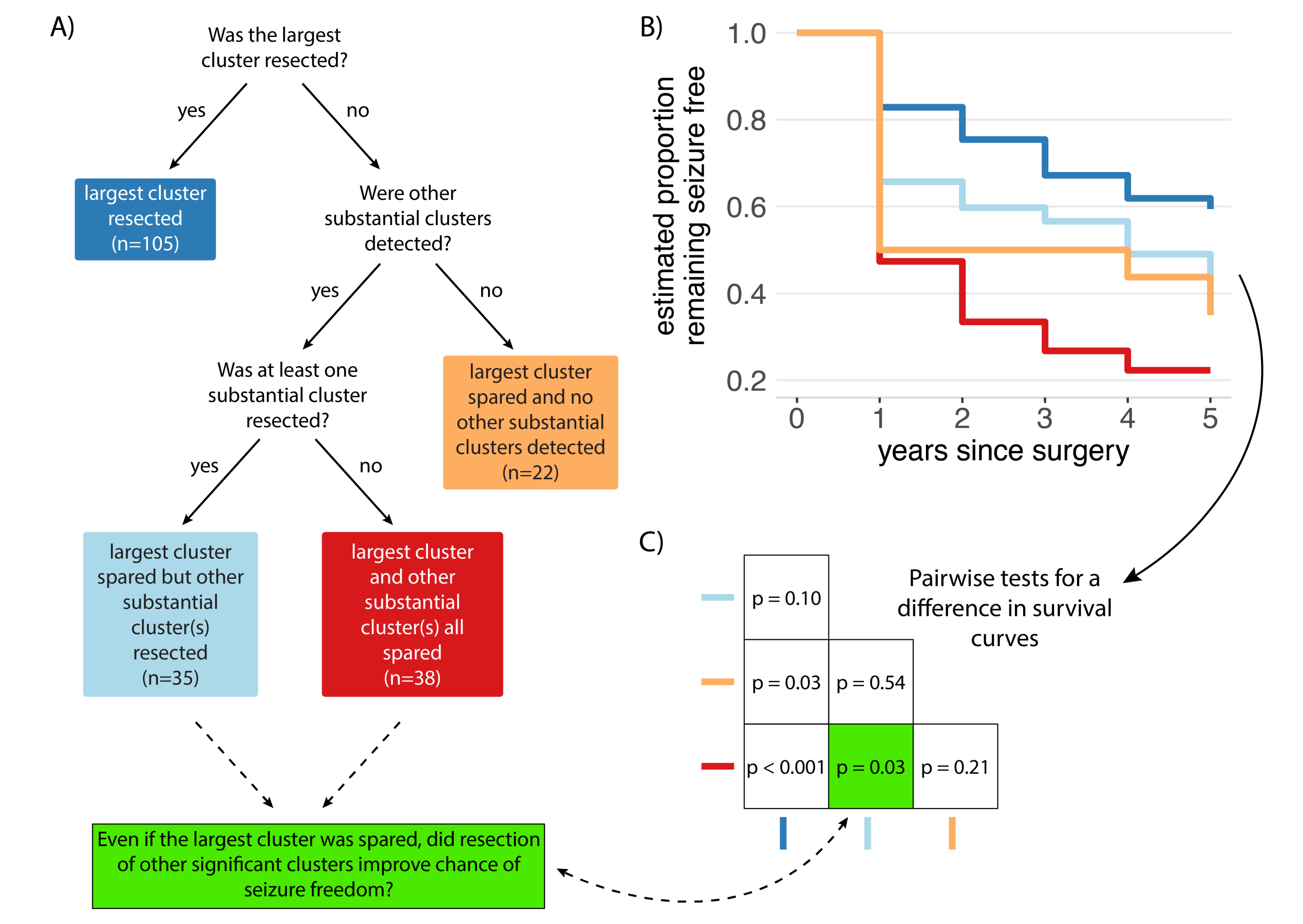

***Figure 5: Resection of other substantial clusters may improve seizure freedom**. A) A decision tree was used to investigate whether, if the largest cluster was spared, the resection of other substantial clusters related to an improved chance of seizure freedom. B) and C) Survival analysis showed that if the largest cluster was spared and other substantial clusters were detected, the resection of these other substantial clusters related to an improved chance of seizure freedom ($p = 0.03$). The best rates of seizure freedom occurred when the largest cluster was resected.*

## Discussion

In this work, we present an approach to identify clusters of abnormalities from diffusion-weighted MRI. We show that resection of these clusters may lead to an improved likelihood of seizure freedom in people with epilepsy. We demonstrated robustness across scanning type with replication in two independent cohorts.

The detection of epileptogenic abnormalities is crucial to improve rates of seizure freedom after surgery for drug-resistant focal epilepsy. People may continue to have seizures after surgery for several reasons, including a) if the correct part of the brain was targeted but resection size was insufficient, b) the wrong part of the brain was targeted or multifocal epileptogenic regions. Our method detects abnormalities in individual voxels, which may delineate a) the extent and b) the location of the epileptogenic zone. Importantly, our results are consistent when using only those subjects who were classified as MRI-negative by clinicians (Supplementary Analysis 3). Crucially, these people also have lower rates of seizure freedom[43] and uncertain localisation of the epileptogenic zone is the main reason for not having epilepsy surgery[44]. Another benefit of our approach is the impact on minimally invasive but maximally effective surgeries[45]. If the suspected epileptogenic tissue can be clearly delineated with high precision, then the removal of adjacent healthy brain tissue can be minimised. This fine-grained abnormality detection could also have applications beyond resective surgery, for more localised treatment techniques such as laser interstitial thermal therapy (LITT)[46], targeted gene therapy[47,48] and deep brain stimulation[49]. Regardless of the eventual treatment option, the clear identification of epileptogenic tissue is of huge clinical importance.

We showed that resection of the largest cluster was associated with an improved likelihood of seizure freedom, where resection was defined as any overlap between the largest cluster and the resection mask. Importantly, we also demonstrated that complete resection of the largest cluster was not required for an improved rate of seizure freedom. This may be important if all of the largest cluster cannot be resected due to proximity to eloquent cortex (e.g. Patient 3 in Figure 3A). The risk of a significant deficit resulting from a resection in brain areas associated with motor or language function is another common reason for not having epilepsy surgery[44]. Resecting only the part of the largest cluster not in eloquent cortex may therefore optimise the likelihood of seizure freedom, whilst minimising the risk to crucial functions.

Some people achieved seizure freedom after surgery, despite not having the largest (or any substantial) cluster resected. It is evident that not all areas with abnormal dwMRI signal are epileptogenic. There were people with abnormal clusters in multiple lobes that were not all resected (Figure 4B) and who became seizure free. In addition, epilepsy is a network disorder[50], and epilepsy surgery has a significant impact on the wider structural connectome[51]. The disconnection of critical white matter tracts may have widespread effects. Resection may have sufficient impact on the epileptogenic network to prevent further seizures even if the the largest (or any substantial) cluster were not resected. Future work could extend our analyses to consider the *hubness* of individual voxels, in addition to the abnormality, to provide additional insight[52], since network properties have previously been shown to relate to outcome[16,18,51,53,54].

Mean diffusivity (MD) quantifies the average diffusion in all directions in each voxel in the brain. Unrestricted diffusion, as seen in free water, results in higher MD values than in tightly packed neurons. Increased white matter MD is often reported in epilepsy[4], and is thought to reflect myelin disruption and increased extracellular space[55,56]. Our results are similarly driven by MD increases (Supplementary Analysis 9). In our main analysis, we do not restrict ourselves to solely white matter voxels, since grey matter is clearly crucial in epilepsy[57,58] (see Supplementary Analysis 10 for white matter only analysis). Grey matter diffusion abnormalities may represent a breakdown in cellular microstructure and have been investigated for their use as biomarkers in other neurological disorders[59]. Future work will investigate the relative predictive ability of abnormality clusters derived from multi-compartment metrics, which may relate more directly to underlying tissue microstructure than the traditional diffusion tensor metrics used in this study[21,60,61].

This study has limitations. The abnormalities that we calculate do not account for age and sex due to the computational and technical infeasibility of regressing out these covariates in every voxel. Our results are promising, given the relationship between diffusion tensor values and both age and sex in health[62,63]. New methods to account for these covariates should only improve the accuracy of calculated abnormalities. In addition, abnormalities were computed by comparison to two relatively small cohorts of controls. To further strengthen this work in the future, normative models of diffusion-weighted MRI should be developed, as they have in other modalities[64–72]. These normative models should be trained on a large number of controls and can act as a comprehensive healthy baseline against which abnormalities can be calculated.

In summary, we present an approach for the identification of focal brain abnormalities from dwMRI. We show retrospectively that, following epilepsy surgery, the likelihood of seizure freedom may be improved by the resection of clusters of dMRI abnormalities. The identification and resection of these abnormalities prospectively has the potential to inform clinical decision making and improve outcomes. Mechanistically, we interpret these abnormal clusters as potential network disruptions, and as such, make the important clinical conclusion even partial resection can significantly increase the chance of seizure freedom.

## Acknowledgements

We thank members of the Computational Neurology, Neuroscience & Psychiatry Lab (www.cnnp-lab.com) for discussions on the analysis and manuscript. J.J.H. is supported by the Centre for Doctoral Training in Cloud Computing for Big Data (EP/L015358/1). P.N.T. is supported by a UKRI Future Leaders Fellowship (MR/T04294X/1). Y.W. gratefully acknowledges funding from Wellcome Trust (208940/Z/17/Z) and is supported by a UKRI Future Leaders Fellowship (MR/V026569/1). G.W. was supported by the MRC (G0802012, MR/M00841X/1). J.D. is grateful to Wellcome Trust (WT106882) and Epilepsy Research UK. J.D. and J.dT. are supported by the NIHR UCLH/UCL Biomedical Research Centre. We are grateful to the Epilepsy Society for supporting the Epilepsy Society MRI scanner. This work was supported by the National Institute for Health Research University College London Hospitals Biomedical Research Centre. The authors acknowledge the facilities and scientific and technical assistance of the National Imaging Facility, a National Collaborative Research Infrastructure Strategy (NCRIS) capability, at the Centre for Microscopy, Characterisation, and Analysis, the University of Western Australia.